\documentstyle[11pt,epsfig]{article}

\def\laq{~\raise 0.4ex\hbox{$<$}\kern -0.8em\lower 0.62
ex\hbox{$\sim$}~}
\def\gaq{~\raise 0.4ex\hbox{$>$}\kern -0.7em\lower 0.62
ex\hbox{$\sim$}~}

\begin{document}

\begin{titlepage}
\begin{flushright}
\end{flushright}
\vspace*{1.5 cm}

\begin{center}
\huge{Uniform gradient expansions}
\vskip 1cm
\large{Massimo Giovannini\footnote{e-mail address: massimo.giovannini@cern.ch}}
\vskip 0.5cm
{\it   Department of Physics, Theory Division, CERN, 1211 Geneva 23, Switzerland}
\vskip 0.5cm
{\it  INFN, Section of Milan-Bicocca, 20126 Milan, Italy}
\vskip 1cm

\begin{abstract}
Cosmological singularities are often discussed by means of a gradient expansion
 that can also describe, during a quasi-de Sitter phase,
the progressive suppression of curvature inhomogeneities. 
While the inflationary event horizon is being formed the two mentioned regimes coexist and  
 a uniform expansion can be conceived and applied to the evolution of spatial gradients across the protoinflationary 
boundary. It is argued that conventional arguments addressing the preinflationary initial conditions 
are necessary but generally not sufficient to guarantee a homogeneous onset of the conventional inflationary stage.
\end{abstract}
\end{center}
\end{titlepage}

\newpage
The dynamical approach to the cosmological singularity has been historically 
 investigated in terms of an expansion in spatial gradients of the geometry \cite{bel1,bel2} (see also \cite{LL}).
Denoting with $t$ the cosmic time coordinate, the gradient expansion in the proximity of the big-bang singularity is formally defined in the limit $t\to 0$ where the spatial  gradients turn out to be subdominant in comparison with the extrinsic curvature. This important 
observation implies that close to the singularity the geometry may be highly anisotropic but rather homogeneous \cite{bel1,bel2}. 
As soon as an inflationary event horizon is formed, the physical rationale for a complementary gradient expansion emerges in the limit $t\to \infty$ \cite{star,salope,tomita}. This idea is applied, for instance, when arguing in favour of the so-called cosmic no-hair conjecture stipulating that in conventional inflationary models any finite portion of the event horizon gradually loses the memory of an initially imposed anisotropy or inhomogeneity so that the metric attains the observed regularity regardless of the initial boundary conditions (see Ref. \cite{hoyle} for this formulation of the conjecture and also 
Refs. \cite{zel1,misner} for some other early contributions). According to the standard lore, one of the central motivations of the whole inflationary paradigm (see e.g. \cite{wein,inf}) is to wash out primeval anisotropies in the expansion right after the formation of the inflationary event horizon 
(see, however, Ref. \cite{barrow} for a critical perspective on the limitations of the no-hair conjecture).

Over a time scale 
${\mathcal O}(t_{*})$ corresponding to the formation of the inflationary event horizon, it is therefore plausible to analyze the space-time geometry not only in terms of a backward gradient expansion (valid for $t < t_{*}$)  and but also by means of a forward expansion (applicable for $t > t_{*}$). 
In both regimes, following  the synchronous Adler-Deser-Misner parametrization \cite{ADM1} the four-dimensional metric tensor can be decomposed as $g_{00} =1$, $g_{ij}= - \gamma_{ij}(\vec{x},t)$ and $g_{0i} =0$. The six independent entries of $\gamma_{ij}(\vec{x},t)$ can be expanded as: 
\begin{equation}
\gamma_{ij}(\vec{x},t) = a^2(t) \biggl[\alpha_{ij}(\vec{x}) + \beta_{ij}(\vec{x},t) + \, .\,.\,.\biggr],
\label{grad1}
\end{equation}
where $\beta_{ij}(\vec{x},t)$ contains two spatial gradients and the ellipses stand for the higher terms in the expansion containing 
a progressively larger (even) number of spatial gradients. Once the inhomogeneous seed metric $\alpha_{ij}(\vec{x})$ is assigned,
the Einstein equations together with the equations of the sources determine $\beta_{ij}(\vec{x},t)$ whose explicit form can
always be parametrized in terms of two dimensionless functions that shall be conventionally called $f(t)$ and $g(t)$:
\begin{equation}
\beta_{i}^{j}(\vec{x},t) = f(t) \frac{{\mathcal P}_{i}^{j}(\vec{x})}{H_{*}^2} + g(t) \frac{{\mathcal P}(\vec{x})}{H_{*}^2}\, \delta_{i}^{j},
\label{ans1}
\end{equation}
where ${\mathcal P}_{i}^{j}(\vec{x}) = a^2(t) {\mathcal R}_{i}^{j}(\vec{x},t)$ is expressed in units of $H_{*}^2 \simeq t_{*}^{-2}$ and ${\mathcal R}_{ij}(\vec{x},t)$ denotes the three-dimensional Ricci tensor.  The evolution of $f(t)$ and $g(t)$ depends, in its turn, on the zeroth-order solution.  If the expansion 
of Eqs. (\ref{grad1}) and (\ref{ans1}) can be safely applied,  the Universe is already quasi-homogeneous in a time interval centered around $t_{*}$ and this 
will be our first assumption on the process describing the formation of the event horizon. Secondly we shall posit that, for $t < t_{*}$, the zeroth-order solution expands in a decelerated manner while it inflates for $t > t_{*}$:  roughly speaking this assumption implies that $t_{*}$ can be identified with the ptotoinflationary boundary. We shall finally admit that the zeroth-order 
evolution of the extrinsic curvature is continuous and monotonic: this last assumption can be relaxed but it is nonetheless realized in the explicit toy examples illustrated hereunder.
Are the three aforementioned assumptions sufficient to guarantee 
that the spatial gradients of the geometry are exponentially suppressed for $t \gg t_{*}$? Are they compatible with the asymptotic 
suppression of the spatial gradients during the quasi-de Sitter stage?
The two previous questions can be approached within a uniform gradient expansion holding across the protoinflationary boundary.

While the fully inhomogeneous inflationary initial conditions represent a rather complicated topic whose proper formulation is beyond the scopes of this 
paper, in what follows we shall content ourselves with the conventional lore in a system where the inflaton field 
$\varphi$ evolves under the action of its own potential $W(\varphi)$ and in the presence of spatial inhomogeneities (characterized by the 
three-dimensional Ricci scalar ${\mathcal R}$);  to account for a possible decelerated behaviour in the preinflationary epoch, we shall also include 
the contribution of an ambient relativistic fluid whose energy density will be denoted by $\rho$.  When the various components of the system are all in equipartition we approximately have\footnote{The Planck mass will be defined as $\overline{M}_{\mathrm{P}}= 1/\sqrt{8 \pi G}$ where $G$ is the Newton constant; the Planck length, in these natural units, is just the inverse of $\overline{M}_{\mathrm{P}}$, i.e. $\ell_{\mathrm{P}} = \sqrt{8 \pi G}$.}:
\begin{equation}
\dot{\varphi}^2 \simeq W(\varphi) \simeq \rho \simeq {\mathcal R} \overline{M}_{\mathrm{P}}^2,
\label{lim1}
\end{equation}
where the overdot denotes a derivation with respect to the cosmic time coordinate $t$. 
Since the kinetic energy, the spatial curvature and the fluid energy density are all diluted faster than $W(\varphi)$, 
Eq. (\ref{lim1}) implies, in the conventional lore, that the Universe inflates before becoming inhomogeneous; this conclusion holds provided
 the background was expanding prior to $t_{*}$.  A successful  inflationary dynamics can also be realized in other situations compatible with Eq. (\ref{lim1}) like, for instance,  
$W(\varphi) \gg\, {\dot{\varphi}}^2 \simeq\, \rho \simeq \,{\mathcal R} \overline{M}_{\mathrm{P}}^2$: also in this case all the 
components of the energy-momentum tensor will quickly disappear and the potential will dominate even faster than in the case of Eq. (\ref{lim1}).
Conversely, if the approximate equipartition of Eq. (\ref{lim1}) is violated,  the typical scale of the potential gets much smaller than the other components of the system:  various inverted hierarchies can be envisaged and they turn out to be particularly relevant in the case of plateau-like potentials \cite{inf}. For instance it can happen that 
${\dot{\varphi}}^2\, \simeq \rho\, \simeq {\mathcal R} \overline{M}_{\mathrm{P}}^2\, \gg W$:
in this case the kinetic energy is diluted more rapidly than the other terms and, after few efolds,  the spatial gradients 
contained in ${\mathcal R}$ dominate the evolution of the sources while the potential is still too small to play any role so that the Universe 
fails to inflate\footnote{Other potentially dangerous hierarchies are, for instance, $\dot{\varphi}^2 \gg W \gg {\mathcal R} \overline{M}_{\mathrm{P}}^2 \simeq \rho$ and  ${\dot{\varphi}}^2 \gg W \simeq {\mathcal R} \overline{M}_{\mathrm{P}}^2\gg \rho$.}. For the present ends what matters is not the likelihood of inflation (or its naturalness) given a generic set of initial data but just the observation 
that Eq. (\ref{lim1}) and its descendants are based on the scaling  properties of the various components 
of the total energy-momentum tensor under the implicit assumption that the geometry is already expanding. We shall therefore grant that the initial 
stages of the inflationary phase are continuously preceded by an epoch where the geometry expands in a decelerated manner and study, in this 
standard framework, the evolution of the spatial gradients.

Within the conventional formulation of the inflationary initial conditions it can be naively expected that $f(t)$ and $g(t)$ will be going 
to zero as a power (for $t< t_{*}$) and 
quasi-exponentially  (for $t >t_{*}$). The governing equations of the system imply that the evolution of $g(t)$ depends directly on the sources (see below, Eqs. (\ref{greq1a})--(\ref{greq3b})) while in the case of of $f(t)$ the evolution reads:
\begin{equation}
 \ddot{f} + 3 H \dot{f} + 2 H_{*}^2 \biggl(\frac{a_{*}}{a}\biggr)^2 =0,\qquad H= \frac{\dot{a}}{a}.
 \label{bas1}
 \end{equation}
Introducing the initial integration time $t_{i}$, the solution of Eq. (\ref{bas1}) depends on $f_{i}= f(t_{i})$ and $\dot{f}_{i} = \dot{f}(t_{i})$  and can be written as:
 \begin{eqnarray}
\dot{f}(t) &=& \dot{f}_{i} \,\biggl(\frac{a_{i}}{a}\biggr)^{3} - 2 H_{*}^2 \biggl(\frac{a_{*}}{a}\biggr)^3 \, \int_{t_{i}}^{t} \frac{a(t^{\prime})}{a_{*}} \,dt^{\prime},
\nonumber\\
f(t) &=& f_{i} + a_{*}^3\,  \dot{f}_{i}\,\int_{t_{i}}^{t} \,\frac{d \, t^{\prime}}{a^3(t^{\prime})} - 2 H_{*}^2 a_{*}^2 \int_{t_{i}}^{t} 
\frac{ d t^{\prime}}{a^3(t^{\prime})} \int_{t_{i}}^{t^{\prime}} a(t^{\prime\prime})\, d t^{\prime\prime}.
\label{sol1}
\end{eqnarray}
The explicit form of $a(t)$ is obtainable by solving  the zeroth-order in the gradient expansion but let us just assume that $\ddot{a}  < 0$ and $\dot{a} >0$ 
for $t < t_{*}$. Such a functional behaviour is realized, for instance\footnote{Note, incidentally, that if the preinflationary background is dominated by a perfect fluid 
with constant barotropic index $w$, then $\delta = 3 (w+1)/2$; conversely if the preinflationary background is dominated by the kinetic 
energy of the inflaton (and the ambient fluid is absent) $\delta \to 3$. }, 
when $ a(t) \sim a_{*}(t/t_{*})^{1/\delta}$ provided  $1 < \delta \leq 3$.  For $t > t_{*}$ we posit instead that 
$\ddot{a} >0$ and $\dot{a} >0$ and the conventional inflationary dynamics implies $\epsilon = - \dot{H}/H^2 \laq 1$.
Under the conditions expressed by  Eq. (\ref{lim1}) the solution of  Eq. (\ref{bas1}) in the two asymptotic limits, naively implies\footnote{If regarded 
in cosmic time, the requirements of Eq. (\ref{expect}) translate in an approximate interpolating form of $f(t)$ that could be written, up to slow roll corrections, as $f(t) \simeq (t/t_{*})^{2(\delta-1)/\delta +1}/[ e^{2 H_{*} t} -1]$. As we shall demonstrate, this plausible guess, implying $\dot{f}(t) \simeq 0$ for $t\simeq t_{*}$, is not supported by the explicit dynamics of the spatial gradients.} :
\begin{equation}
 \lim_{a\ll a_{*}} f(a) \,\to \biggl(\frac{a}{a_{*}}\biggr)^{2(\delta-1)}, \qquad  \lim_{a\gg a_{*}} f(a)\, \to \biggl(\frac{a}{a_{*}}\biggr)^{-2 + 2 \epsilon}.
\label{expect}
\end{equation}
Not surprisingly, Eq. (\ref{expect}) is consistent with the results separately obtainable in the two limits (see, e.g. \cite{bel1,bel2} and \cite{star,salope,tomita}) but what matters here is that such a condition seems to demand the existence of an extremum for $a\sim {\mathcal O}(a_{*})$ or $t\simeq {\mathcal O}(t_{*})$.
According to Eq. (\ref{sol1}) the existence of a maximum would imply that $|\dot{f}(t)| \to 0$ for $t\simeq t_{*}$, where the absolute value accounts for the possibility of negative values of $f(t)$. The vanishing of $\dot{f}(t)$ can occur either for finite cosmic time (but then we must have that $\dot{f}_{i} \neq 0$) or 
asymptotically for $t\gg t_{*}$. The choice  $\dot{f}_{i} \neq 0$ causes the presence of divergent term in the limit $t \ll t_{*}$ and this 
clashes with the possibility of imposing quasi-homogeneous initial conditions in the preinflationary phase, as conventionally assumed.
According to this argument, what can happen, at most is  $|\dot{f}| \to 0$ for $t\gg t_{*}$; if this is the case 
the gradients will not be asymptotically suppressed but $f(t)$ will rather reach a constant value.
Thus the smooth and monotonic evolution of the extrinsic curvature 
across the protoinflationary transition does not seem sufficient to guarantee that the spatial gradients will be exponentially 
suppressed during the fully developed inflationary phase. The simplistic way of reasoning pursued in this paragraph assumes, without proof,  
a certain behaviour of the scale factor. In what follows we shall then focus the attention 
to the full zeroth-order and first-order solutions in the case when the extrinsic curvature interpolates 
between a decelerated regime and an accelerated evolution in the vicinity of $t_{*}$.

We are now ready to consider the general system of equations: 
separating the extrinsic curvature ($K_{ij}= - \dot{\gamma}_{ij}/2$) from the contribution of the intrinsic curvature 
(${\mathcal R}_{ij}$), the $(00)$ and $(0i)$ components of the contracted Einstein equations read:
\begin{eqnarray}
&&\dot{K} - {\rm Tr} K^2 = \ell_{\mathrm{P}}^2 \,\biggl[ \frac{(3 p+ \rho)}{2} + (p + \rho) u^2 + \dot{\varphi}^2 - W(\varphi) \biggr],
\label{HAM0}\\
&& \nabla_{i} K - \nabla_{k} K^{k}_{i} =  \ell_{\mathrm{P}}^2 \biggl[ u_{i}  \sqrt{ 1 + u^2} (p + \rho) + \dot{\varphi} \partial_{i} \varphi\biggr],
\label{MOMi}
\end{eqnarray}
where $u_{i}$ is the total velocity of the fluid and $u_{0} = u^{0} = \sqrt{1 + u^2}$ with $u^2 = \gamma^{ij} u_{i} u_{j}$. 
In Eq. (\ref{HAM0}) the compact notation ${\rm Tr} K^2 = K_{i}^{j} \, K_{j}^{i}$  has been used; $\nabla_{i}$ denotes the covariant derivative defined 
with respect to the metric $\gamma_{ij}$. Finally, the $(ij)$ component of the contracted Einstein equations reads:
\begin{equation}
\dot{K}_{i}^{j} - K \,K_{i}^{j} - {\mathcal R}_{i}^{j} = - \ell_{\mathrm{P}}^2 \biggl[ (p+ \rho) u_{i} u^{j} 
+\partial_{i} \varphi \partial^{j} \varphi  -  \frac{\rho - p}{2} + W(\varphi) \biggl] \delta_{i}^{j}.
\label{E2ij}
\end{equation}
Inserting Eq. (\ref{grad1}) into Eqs. (\ref{HAM0}) and (\ref{E2ij}), to zeroth order we shall have:
\begin{eqnarray}
&& 6 \overline{M}_{\mathrm{P}}^2 (\dot{H} + H^2) +\rho^{(0)} + 3 p^{(0)} + 2 \,\dot{\varphi}^{(0)\, 2} - 2\,W[\varphi^{(0)}]=0,
\label{FR1}\\
&&  2 \overline{M}_{\mathrm{P}}^2 ( \dot{H} + 3 H^2) - \rho^{(0)} + p^{(0)} - 2\,W[\varphi^{(0)}] =0,
\label{FR2}
\end{eqnarray}
where $p(\vec{x},t)$ is the pressure of the fluid and the superscript denotes the order of the expansion of the sources:
\begin{equation}
p(\vec{x},t) = p^{(0)}(t) + p^{(1)}(\vec{x},t), \qquad \rho(\vec{x},t) = \rho^{(0)}(t) + \rho^{(1)}(\vec{x},t), \qquad \varphi(\vec{x},t) = \varphi^{(0)}(t) + \varphi^{(1)}(\vec{x},t).
\label{grad2}
\end{equation}
  Equations (\ref{FR1}) and (\ref{FR2}) are supplemented by the zeroth-order forms of the continuity equations 
  $\dot{\rho}^{(0)} + 3 H ( \rho^{(0)} + p^{(0)}) =0$ and  of the Klein-Gordon equation $\ddot{\varphi}^{(0)}  + 3 H  \dot{\varphi}^{(0)} + W_{,\,\varphi}[\varphi^{(0)}]=0$. To first-order Eqs. (\ref{HAM0}), (\ref{MOMi}) and (\ref{E2ij}) imply respectively:
\begin{eqnarray}
&& \ddot{\beta} + 2 H \dot{\beta} + \ell_{\mathrm{P}}^2 \biggl(\rho^{(1)} + 3 p^{(1)} + 4 \dot{\varphi}^{(0)} \dot{\varphi}^{(1)} - 2 W_{,\,\varphi}[\varphi^{(0)}] \varphi^{(1)}\biggr) =0,
\label{greq1}\\
&&   \nabla_{j} \dot{\beta}^{j}_{i} - \nabla_{i} \dot{\beta} =  2 \ell_{\mathrm{P}}^2 \biggl[ ( p^{(0)} + \rho^{(0)}) u_{i} + \dot{\varphi}^{(0)} \partial_{i} \varphi^{(1)}\biggr],
\label{greq2}\\
&& \ddot{\beta}_{i}^{j} + 3 H \dot{\beta}_{i}^{j} + \frac{2}{a^2} {\mathcal P}_{i}^{j}  + H \dot{\beta} \delta_{i}^{j} = \ell_{\mathrm{P}}^2 \biggl( \rho^{(1)} - p^{(1)} + 2 W_{,\,\varphi}[\varphi^{(0)}]\, \varphi^{(1)}\biggr )\delta_{i}^{j},
\label{greq3}
\end{eqnarray}
where $W_{,\,\varphi}[\varphi^{(0)}]$ denotes the first derivative of the potential with respect to $\varphi$ evaluated for $\varphi=\varphi^{(0)}$.
Even though the evolution equations of the sources are consequences of the previous equations (exactly as their zeroth-order counterparts) it is 
useful to write them in some detail:
\begin{eqnarray}
&& [ p^{(0)} + \rho^{(0)}] \dot{u}_{i} + \dot{p}^{(0)} u_{i} = \partial_{i} p^{(1)},
\label{greq4}\\
&& \dot{\rho}^{(1)} + \frac{\dot{\beta}}{2} [ p^{(0)} + \rho^{(0)}]  + 3 H  [ p^{(1)} + \rho^{(1)}]  =0,
\label{greq5}\\
&& \ddot{\varphi}^{(1)} + 3 H \dot{\varphi}^{(1)} + \frac{\dot{\beta}}{2} \dot{\varphi}^{(1)} + W_{,\varphi\varphi}[\varphi^{(0)}] \varphi^{(1)} =0.
\label{greq6}
\end{eqnarray}

As anticipated 
we shall now focus the attention on those explicit solutions exhibiting a monotonic behaviour of the extrinsic curvature. 
In the absence of fluid sources a sound zeroth-order solution satisfying Eqs. (\ref{FR1})--(\ref{FR2}) is:
\begin{equation}
a(t) = a_{*} \sinh^{1/3}{(3 \, H_{*} t)}, \qquad \dot{\varphi}^{(0)\,2}(t) = \frac{6 \overline{M}_{\mathrm{P}}^2 H_{*}^2}{ \sinh^2{(3\, H_{*}\, t)}}.
\label{puresc3}
\end{equation}
In the limit $t \gg t_{*} \simeq H_{*}^{-1}/3$ the Universe inflates and $ \dot{\varphi}^{(0)} \to 0$. In the limit $t \ll t_{*}$ the solution is instead
decelerated going asymptotically as $(t/t_{*})^{1/3}$.  Sticking to the case of a constant barotropic index\footnote{This choice implies 
 the absence of non-adiabatic fluctuations in the system. This property will translate, ultimately, in a simpler form 
 of the first-order solution. More general situations can be considered but are not central to the present discussion.}, the full solution of Eqs. (\ref{FR1}) and (\ref{FR2}) (and of the corresponding equations of the sources) equations  can be expressed as:
 \begin{equation}
a(t) = a_{*} \biggl[ \sinh{(\delta\, H_{*}\, t)}\biggr]^{1/\delta}, \qquad 
 \varphi(t) = \varphi_{0} \pm \sqrt{\frac{2}{\beta}} \,\overline{M}_{\mathrm{P}}\, \sqrt{1 - \Omega_{*}} \,\ln{\biggl[\tanh{\biggl( \frac{\beta H_{*} t}{2}\biggr)}\biggr]}, 
\label{SOL1}
\end{equation}
where $\delta = 3 ( w + 1)/2$. In Eq. (\ref{SOL1})
 we defined the parameter $\Omega_{*}= \rho_{*}/(3 H_{*}^2 \overline{M}_{\mathrm{P}}^2)$ the critical fraction of the ambient fluid 
at the moment of formation of the event horizon; recall, furthermore, that 
$\rho(t)= \rho_{*} (a_{*}/a)^{2 \delta}$. The inflaton potential can be written in this case as:  
\begin{equation} 
W(\varphi) = 3 H_{*}^2 \,\overline{M}_{\mathrm{P}}^2 + 
\frac{3}{2} (1 - w) H_{*}^2 \overline{M}_{\mathrm{P}}^2 (1 - \Omega_{*}) \sinh^2{\biggl[\sqrt{\frac{\beta}{2}} \frac{(\varphi - \varphi_{0})}{(1 - \Omega_{*}) \overline{M}_{\mathrm{P}}}\biggr]}.
\label{SOL4}
\end{equation}
As anticipated, the solution 
satisfies the boundary conditions characterizing the protoinflationary 
transition. In particular for $\delta H_{*} t < 1$ the solution is decelerated and from Eq. (\ref{SOL1}) we have
$a(t) \simeq a_{*} ( \delta H_{*} t)^{1/\delta}$ where $H_{*} \simeq 1/(\delta\, t_{*}) =2/[3 ( w+ 1) t_{*}]$.
In the opposite limit (i.e. $\delta H_{*} t \gg 1$) the solution is accelerated with $H(t) \simeq H_{*}$. In the case 
$\delta \to 3$ (i.e. $w\to 1$) and $\Omega_{*} \to 0$, Eq. (\ref{SOL1}) 
formally gives back Eq. (\ref{puresc3}).

We are now ready to discuss the evolution of $f(t)$ and $g(t)$. Inserting Eq. (\ref{ans1}) into Eqs. (\ref{greq1})--(\ref{greq2}) the explicit form of the corresponding first-order equations is: 
\begin{eqnarray}
&&\ddot{f} + 3 \ddot{g} + 2 H( \dot{f} + 3 \dot{g})  + \ell_{\mathrm{P}}^2  \biggl[( 1 + 3 w) q + 4 \dot{\varphi}^{(0)} \dot{\chi} - 2 W_{,\,\varphi}(\varphi^{(0)}) \chi \biggr]=0,
\label{greq1a}\\
&&  \dot{f} + 4 \dot{g} + 4 \ell_{\mathrm{P}}^2  \biggl[(p^{(0)} + \rho^{(0)}) v +\dot{\varphi}^{(0)} \chi \biggr]=0,
\label{greq2a}
\end{eqnarray}
where, consistently with Eq. (\ref{ans1}) the first-order evolution of the sources has been parametrized as 
$\varphi^{(1)}(\vec{x},t) = \chi(t) {\mathcal P}(\vec{x})/H_{*}^2$,  $\rho^{(1)}(\vec{x}, t) = q(t)  {\mathcal P}(\vec{x})/H_{*}^2$ and 
$u_{i}(\vec{x},t) = v(t) \partial_{i} {\mathcal P}(\vec{x})/H_{*}^2$.
Using the same procedure in the case of Eq. (\ref{greq3}), two separate conditions arise: the first one, as already anticipated, is Eq. (\ref{bas1}) and 
stems from the terms proportional to ${\mathcal P}_{i}^{j}(\vec{x})$; the second condition coming from the coefficient of ${\mathcal P}(\vec{x})$ is:
\begin{equation}
\ddot{g} + 6 H \dot{g} + H \dot{f}  - \ell_{\mathrm{P}}^2  \biggl[( 1 - w) q + 2 W_{,\,\varphi}[\varphi^{(0)}] \chi \biggr]=0,
\label{greq3b}
\end{equation}
and determines the evolution of $g(t)$. Last but not least, the explicit form of  Eqs. (\ref{greq4}) and (\ref{greq6}) becomes:
\begin{equation}
[\rho^{(0)} + p^{(0)}] \dot{v} +  w \dot{\rho}^{(0)} v - w q=0,\qquad \ddot{\chi} + 3 H \dot{\chi} + \frac{(\dot{f} + 3 \dot{g})}{2} \dot{\varphi}^{(0)} + W_{,\,\varphi\varphi}[\varphi^{(0)}] \chi =0.
\label{greq6a}
\end{equation}
The explicit form of Eq. (\ref{greq5}), corresponding to the first-order equation for the energy density of the fluid, 
is directly integrable and the result is $q= -(f+3 g) (p^{(0)} +\rho^{(0)})/2$, assuming $f_{i} = g_{i} = q_{i} =0$.
Equations (\ref{greq1a})--(\ref{greq2a}), (\ref{greq3b}) and (\ref{greq6a}) shall now be solved 
given a set of zeroth-order solutions interpolating between a decelerated stage of expansion and the inflationary phase 
(see e.g. Eqs. (\ref{puresc3}) and (\ref{SOL1})).

Focussing first the attention on the case where the ambient fluid is absent, the solution for $f(t)$ determines  the evolution of $\dot{g}(t)$; via the constraint (\ref{greq1a}), $\chi(t)$ can be eliminated, in the limit $v(t) \to 0$, from Eq. (\ref{greq2a}) (or from Eq. (\ref{greq3b})). The equation for $g(t)$ becomes:
\begin{equation}
 \ddot{g} - 2 \frac{\ddot{\varphi}^{(0)}}{\dot{\varphi}^{(0)}} \dot{g} = \frac{1}{2} \biggl[ \frac{\ddot{\varphi}^{(0)}}{\dot{\varphi}^{(0)}} + H\biggr] \dot{f},
\label{puresc1}
\end{equation}
and its explicit solution is:
\begin{equation}
\dot{g}(t) = \dot{g}_{i} \biggl[\frac{\dot{\varphi}^{(0)}(t)}{\dot{\varphi}^{(0)}_{i}}\biggr]^2+ 
\frac{\dot{\varphi}^{(0)2}(t)}{2}\int_{t_{i}}^{t} \frac{\dot{f}(t^{\prime})}{\dot{\varphi}^{(0)2}(t^{\prime})}
\biggl[ \frac{\ddot{\varphi}^{(0)}(t^{\prime})}{\dot{\varphi}^{(0)}(t^{\prime})} + H(t^{\prime})\biggr] \, d t^{\prime}.
\label{puresc1a}
\end{equation}
As in the case of Eq. (\ref{sol1}),   $|\dot{g}(t_{*})| \to 0$ only if $ \dot{g}_{i} \neq 0$.
We can consequently argue that $f(t)$ and $g(t)$ grow (in absolute value) for $a< a_{*}$ but instead of decreasing for for $a> a_{*}$ they 
reach a constant asymptote without violating the conditions of the gradient expansion. 

Let us finally verify, as a cross-check,  that the standard results of the inflationary gradient expansion are obtainable if the preinflationary 
initial conditions are completely disregarded for $t \ll t_{*}$. More specifically, in the absence of ambient fluid,  the solutions of Eqs. (\ref{bas1}) and (\ref{puresc1}) imply 
 $f(a) \simeq g(a) \simeq \chi(a)/\overline{M}_{\mathrm{P}} \simeq a^{-2 + 2 \epsilon}$ (in the limit $a \gg a_{*}$),
where $\epsilon$ denotes the slow-roll parameter $ \epsilon =  - \dot{H}/H^2$. The coefficients multiplying the power depend on the specific model so, for instance, in the case of purely exponential potentials we shall have $f(a) \sim g(a)\sim \epsilon^2  (a/a_{*})^ {-2 + 2 \epsilon}$ and 
$\chi(a)/\overline{M}_{\mathrm{P}}\sim\epsilon^{5/2} (a/a_{*})^ {-2 + 2 \epsilon}$.
The derived set of equations is also applicable in the absence of scalar field and the only 
contribution is given by the ambient fluid and by its own inhomogeneities. In this case, as a second cross-check, we have that 
the solution of the system for $f_{i}= g_{i} =0$ and $\dot{f}_{i}= \dot{g}_{i} =0$ is given by 
$f(a) = -  4/[( w+1) ( 3 w + 5)] (a/a_{*})^{ 3 w + 1}$ and $g(a) = ( 6 w + 5 - 3 w^2)/[(w+1) (3 w + 5) (9w + 5)] (a/a_{*})^{ 3 w + 1}$.

\begin{figure}[t!]
\begin{center}
\begin{tabular}{|c|c|}
      \hline
      \hbox{\epsfxsize = 6 cm  \epsffile{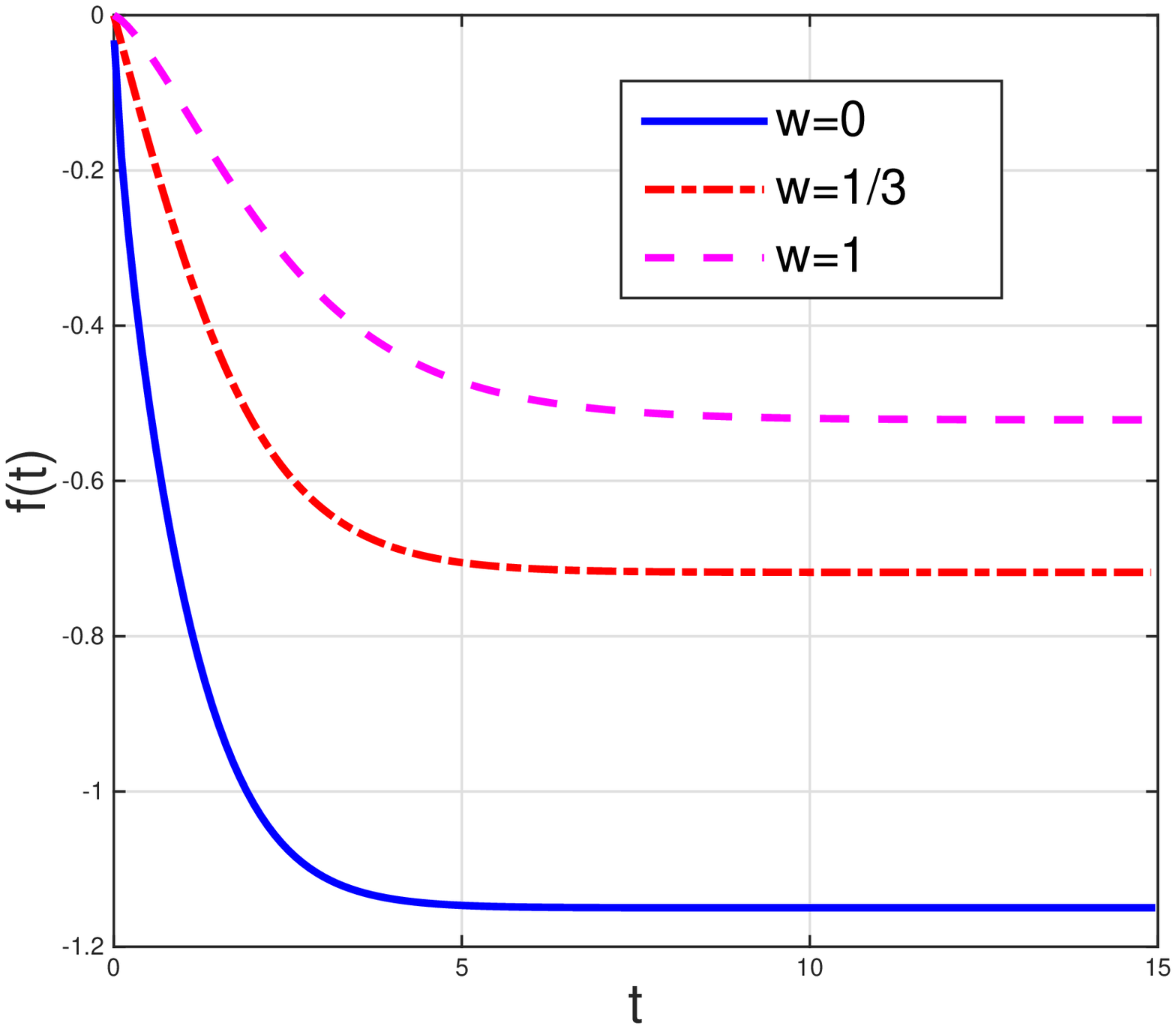}} &
      \hbox{\epsfxsize = 6 cm  \epsffile{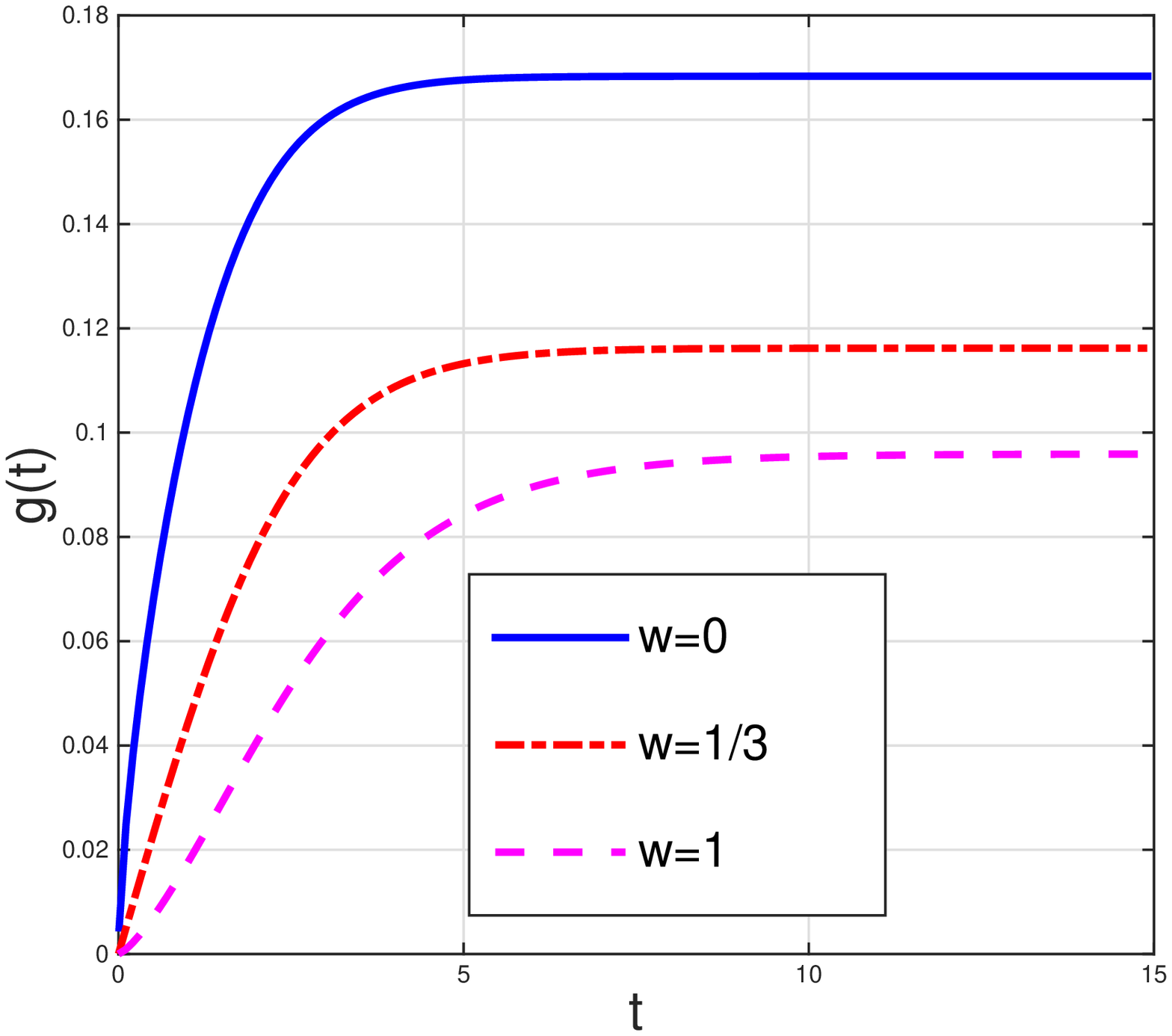}}\\
      \hline
\end{tabular}
\end{center}
\vskip 3mm
\caption[a]{The evolution of $f(t)$ and $g(t)$ when the initial conditions for $t< t_{*}$ solve the zeroth-order and first-order 
system and are reported in Eq. (\ref{SOLAS1}).}
\label{FIGSOL1}
\end{figure}
\begin{figure}[t!]
\begin{center}
\begin{tabular}{|c|c|}
      \hline
      \hbox{\epsfxsize = 6 cm  \epsffile{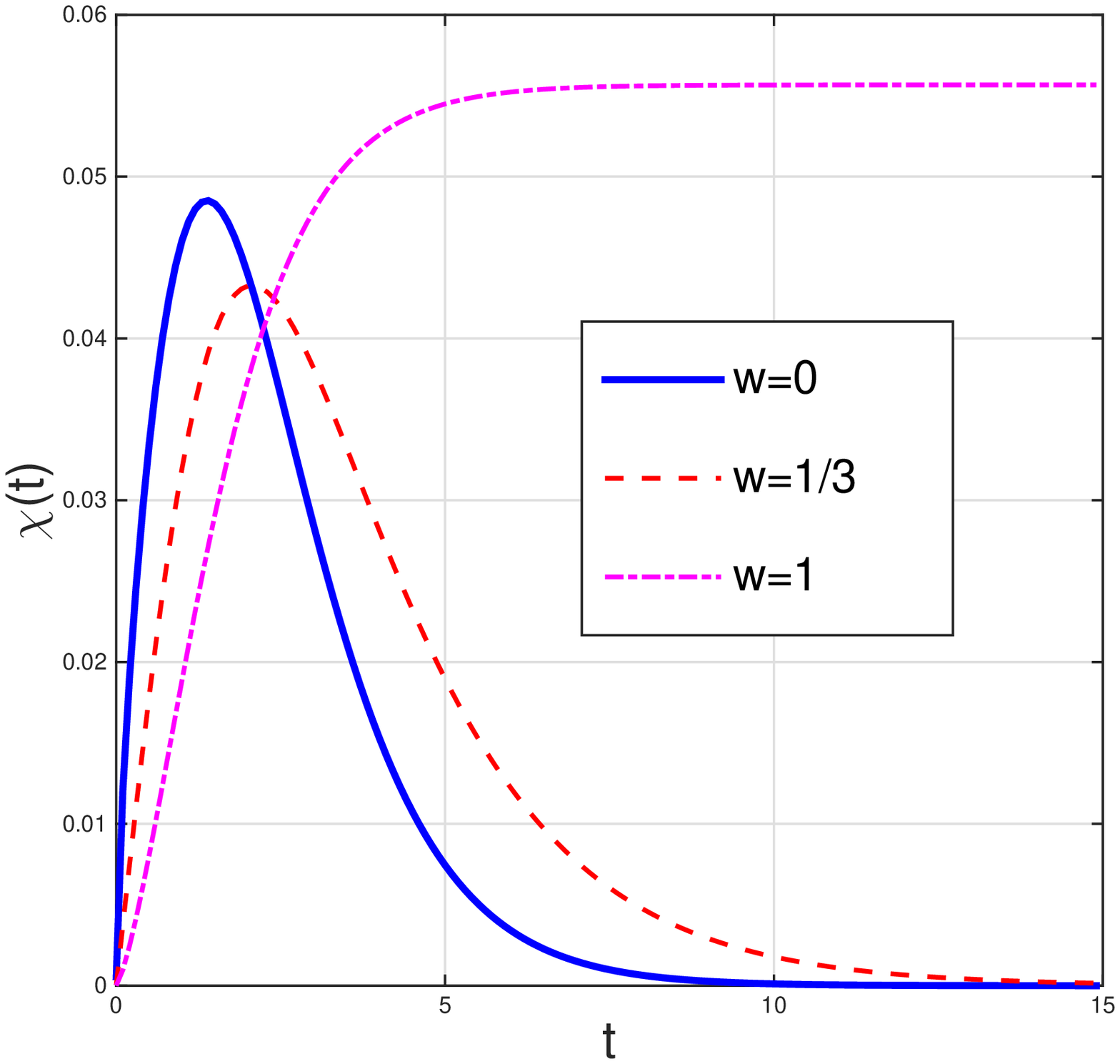}} &
      \hbox{\epsfxsize = 6 cm  \epsffile{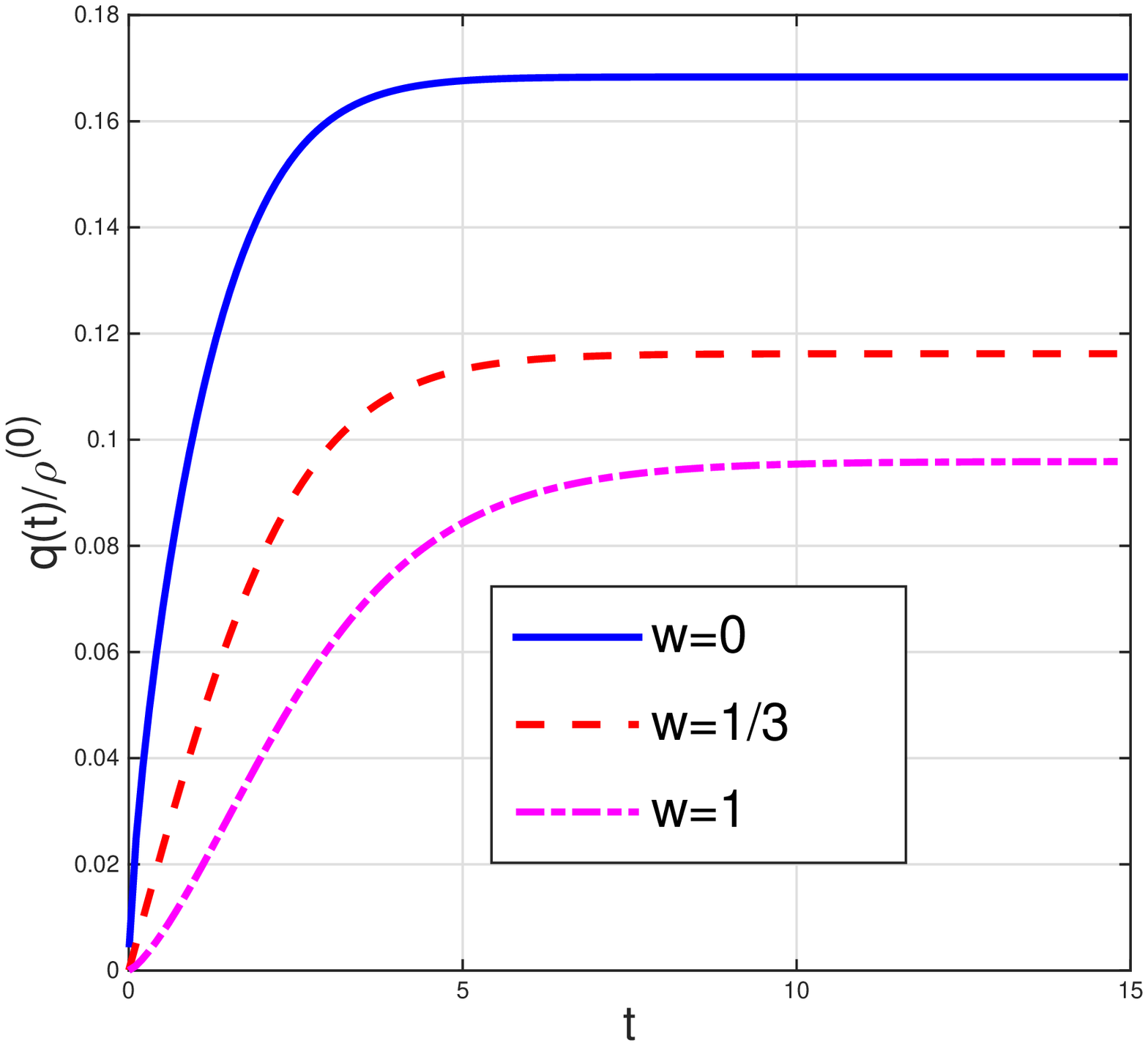}}\\
      \hline
\end{tabular}
\end{center}
\vskip 3mm
\caption[a]{The solution for the first-order evolution of the inflaton and of the ambient fluid for the same initial data of Fig. \ref{FIGSOL1}. The evolution 
of $\chi(t)$ is reported in units $\overline{M}_{\mathrm{P}}= 1$.}
\label{FIGSOL2}
\end{figure}

The most realistic set of zeroth-order solution (see Eq. (\ref{SOL1})) containing simultaneously and ambient fluid and the inflaton
will now be used to solve numerically the corresponding first-order equations; as we shall see the numerical examples corroborate and complete 
the previous analytical arguments.  Given Eq. (\ref{SOL1}), the asymptotic solution of  Eqs. (\ref{greq1a})--(\ref{greq2a}), (\ref{greq3b}) and 
(\ref{greq6a})  for $t< t_{*}$ tuns out to be:
\begin{eqnarray}
&& f(t) = \frac{1}{1-\delta^2} \biggl(\frac{t}{t_{*}}\biggr)^{2(\delta -1)/\delta},\qquad g(t)= {\mathcal A}(\delta)\biggl(\frac{t}{t_{*}}\biggr)^{2(\delta -1)/\delta}, 
\nonumber\\
&&\overline{v}(t) = {\mathcal B}(\delta)\biggl(\frac{t}{t_{*}}\biggr)^{(3\delta -2)/\delta},\,\qquad
\chi(t) = \overline{M}_{\mathrm{P}}\, {\mathcal C}(\delta) \biggl(\frac{t}{t_{*}}\biggr)^{2(\delta -1)/\delta}.
\label{SOLAS1}
\end{eqnarray}
where $\overline{v}(t) = v(t)/t_{*}$ and, as previously mentioned, $\delta = 3(w+1)/2$. 
The three functions ${\mathcal A}(\delta)$, ${\mathcal B}(\delta)$
and ${\mathcal C}(\delta)$ are defined as:
\begin{eqnarray}
&&{\mathcal A}(\delta) = \frac{1}{6} \biggl\{\frac{2}{\delta^2 -1} + \frac{3 \delta-1}{(3 \delta - 2) [ 1 + (\Omega_{*} -1) \delta^2 - 3 \delta \Omega_{*}]}\biggr\},
\nonumber\\
&& {\mathcal B}(\delta) = \frac{ \delta[ ( 11 - 6 \delta) \delta -3]}{ 12 ( 1 + \delta) (3 \delta -2) [ 1 + \delta^2 ( \Omega_{*} -1) - 3 \delta \Omega_{*}]},
\nonumber\\
&& {\mathcal C}(\delta) = \frac{\sqrt{\delta} ( \Omega_{*} -1) (\delta -1)}{2 ( 3 \delta - 2) \sqrt{2 ( 1 - \Omega_{*})} [ 1 + \delta^2 ( \Omega_{*} -1) - 3 \delta \Omega_{*}]}.
\label{SOLAS2}
\end{eqnarray}
Disregarding the preinflationary initial conditions for $f(t)$, $g(t)$ and $\chi(t)$,  the asymptotic solution for $t\gg t_{*}$ can be written as
$(t) = e^{- 2 H_{*} t}$ and $g(t) = - \frac{1}{4} e^{- 2 H_{*} t}$; similarly for $v(t)$ and $\chi(t)$ we have 
$v(t) = w/[4 ( 2 + 3 w) H_{*}] \, e^{- 2 H_{*} t}$, and $\chi(t) = - \overline{M}_{\mathrm{P}}^2 ( \dot{f} + 4 \dot{g})/[4\dot{\varphi}^{(0)}]$.
Equations (\ref{SOLAS1})--(\ref{SOLAS2}) guarantee that  for $t \ll t_{*}$ the solution 
is smooth and quasi-homogeneous. Thus Eqs. (\ref{SOLAS1})--(\ref{SOLAS2}) define the initial conditions of the numerical 
integration for $t_{i} \ll t_{*}$.  Since the zeroth-order solution is characterized by a continuous 
(and monotonic) extrinsic curvature, the first-order equations are integrable across the protinflationary boundary 
and the results of the numerical analysis are reported in Figs. \ref{FIGSOL1} and \ref{FIGSOL2} for different values of the barotropic 
index  and for $\Omega_{*} = 1/10$. The numerical results show that the contribution of the gradients of the geometry is not exponentially 
suppressed but it is asymptotically constant.
In Fig. \ref{FIGSOL1} we illustrate the results in terms of $f(t)$ (left panel) and $g(t)$ (right panel). In Fig. \ref{FIGSOL2}  the evolution 
of $\chi(t)$ and $q(t)/\rho^{(0)}(t)$ is reported. Recall that $q(t)/\rho^{(0)}(t) \propto (f + 3 g)$ (see discussion after Eq. (\ref{greq6a})). 
We did not integrate the constraint of Eq. (\ref{greq2a}) but checked, a posteriori,  that it is obeyed by the 
initial data (\ref{SOLAS2}) and by the full numerical solution to a precision of one part in $10^{6}$.

In summary, the standard conditions for the validity of the quasi-homogeneous gradient expansion together with 
the existence of a smooth evolution of the extrinsic curvature across the protoinflationary boundary do not guarantee 
the exponential suppression of the spatial gradients during the quasi-de Sitter phase that follows a 
preinflationary stage of decelerated expansion. This conclusion has been reached within the first-order in the uniform gradient expansion
by setting the initial conditions of the spatial gradients prior to the formation of the inflationary event horizon. The numerical integration 
 corroborates the analytical expectation and it also suggests that the arguments used to infer the likelihood of inflation on the basis of the scaling properties of the various components of the total energy-momentum tensor are necessary but generally not sufficient to assure
the exponential suppression of the spatial gradients.  Alternatively one may argue that the quasi-homogeneous and quasi-isotropic approximations are not appropriate for describing the formation of the inflationary event horizon. We leave these hypotheses for future investigations.

\end{document}